\documentstyle[pbar2000,11pt,fleqn,epsfig]{article}
\begin{document}
\title{Importance of Future Hyperon \\ Beta Decay Experiments}

\author{Nickolas Solomey \\
Illinois Institute of Technology, Chicago, Illinois 60616}

\maketitle

\begin{abstract}
Recent results from the KTeV experiment at Fermilab using $\Xi^0$
hyperons have enabled a great leap in improving our understanding of
elementary particle physics, especially with the first form-factor
measurement from the semi-leptonic decay $\Xi^0 \rightarrow \Sigma^+
e^- \bar{\nu}$. This decay is a test of whether the standard model
contains all of the needed parameters to fully describe hyperon beta
decay. It was observed for the first time only in 1997 even though its
importance had been explicitly stated in 1961 by the early theories of
the standard model as formulated by N. Cabibbo. We have the ability to
improve this measurement substantially by making the definitive
form-factor measurement with a sample of 30,000 such decays from a
forthcoming experiment, which will either show or rule out the
existence any additional second class weak currents, an obviously
important measurement allowing particle physics to finally put this
question to rest. We also have the ability to make a measurement of
hyperon compositeness by measuring the charged $\Sigma^{\pm}$ beta
decay into $\Lambda^0$, and in addition to search for mass coupling
terms in hyperon beta decays where the muon replaces the electron,
important for determining the $g_3$ and $f_3$ form-factors. These are
the important questions to answer in studying strange baryon decays,
and are reviewed in this article.
\end{abstract}

\section{Review of the Recent KTeV Results}
There are several review articles that summarize the history of hyperon
beta decay \cite{theory1,kaon99solomey}. Here I will remind the reader
of the recent important results and then in the next section elaborate
on how they could be improved upon in future experiments.

The neutral beam of the KTeV experiment was produced by protons from
the Fermilab Tevatron accelerator. It had two components: the rare kaon
decay program, E799, and the search for direct CP violation, E832
\cite{ktev}. Presented here is only a small part of the results from a
neutral hyperon program that had three triggers in the E799 experiment
configuration, with results from both the 1997 and 1999 runs.

\paragraph*{Neutral $K^0$ and hyperon decays:}
The experiment's fiducial decay volume, which starts 90 m downstream of
the target because of the space needed to collimate the neutral beam,
is where most of the particles decay and is also the location of the
sweeping magnets that eliminate the charged particles. The decay volume
from 90 to 160 m from the target was at an ultrahigh vacuum to reduce
interactions and had scintillator ring counters to veto those events
where a particle left the fiducial volume. The spectrometer, consisting
of tracking chambers, an analysis magnet, electromagnetic calorimetry
(CsI) \cite{roodman}, particle identification by transition radiation
detectors (TRD) \cite{solomey}, and a muon counter system with 5 m of
iron filters, directly follows the decay volume. Data were collected
using 16 triggers for two different experimental configurations in 1997
and 1999.

\begin{figure}[h]
\begin{center}
\mbox{\epsfig{file=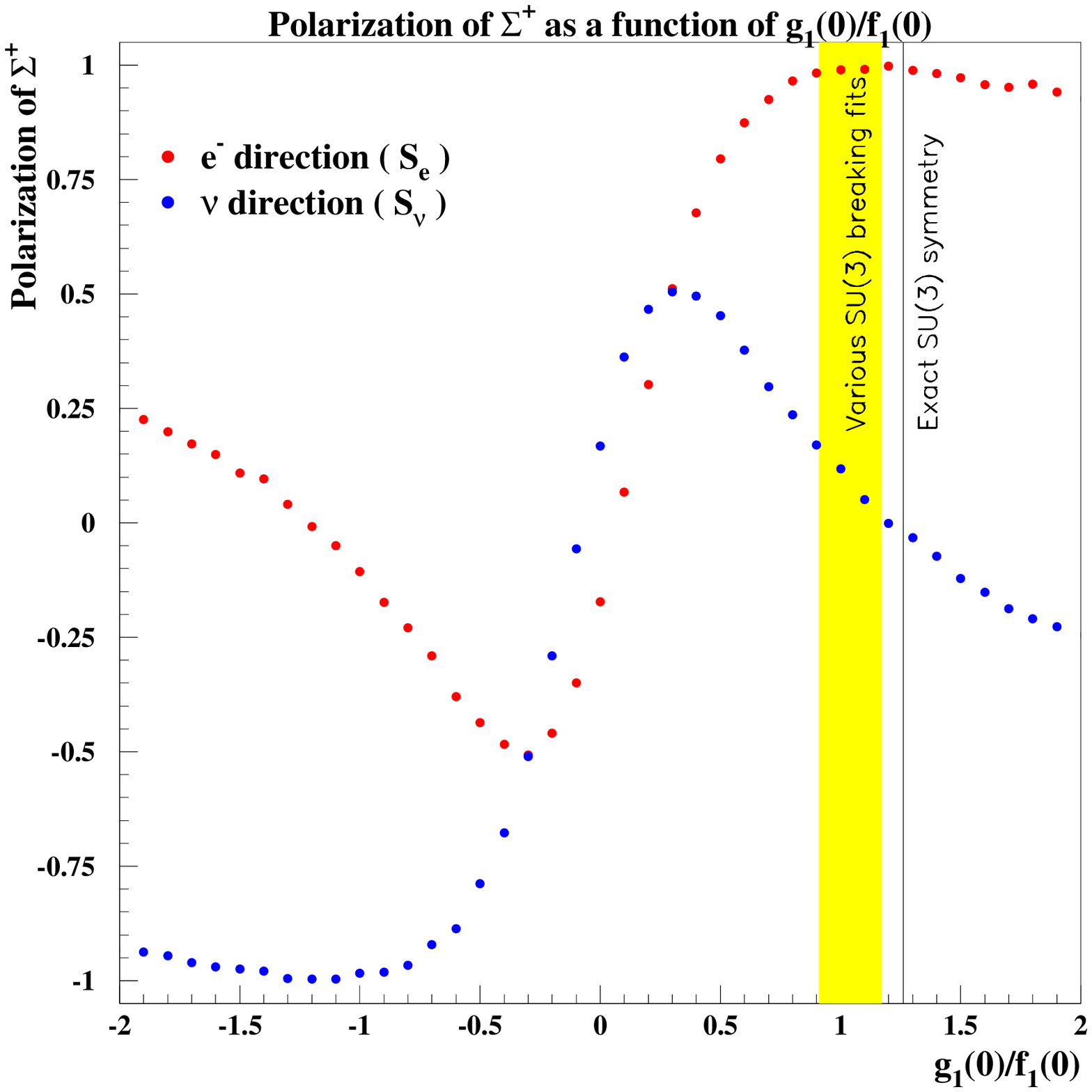,%
height=0.47\linewidth} }
\mbox{\epsfig{file=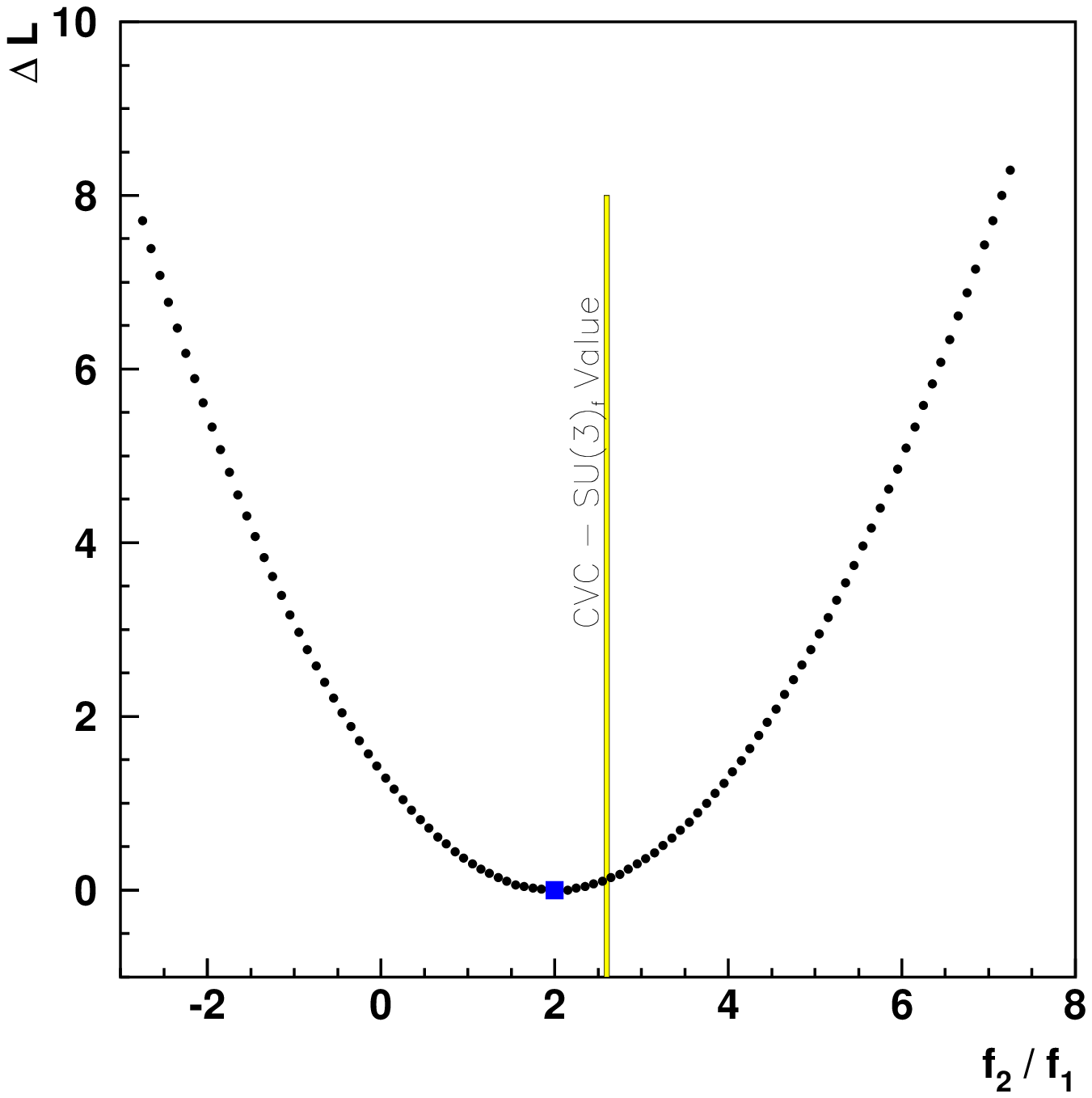,%
height=0.47\linewidth} }
\end{center}
\caption{Theoretical predictions for (left) $\Sigma^+$ polarization for
values of $g_1/f_1$ with SU(3) prediction and various SU(3) symmetry
breaking models indicated by vertical lines, and (right) the $f_2 /f_1$
form-factor, fits from the electron energy spectrum with the SU(3)
prediction.} \label{ff}
\end{figure}

\begin{figure}[h]
\begin{center}
\mbox{
\epsfig{file=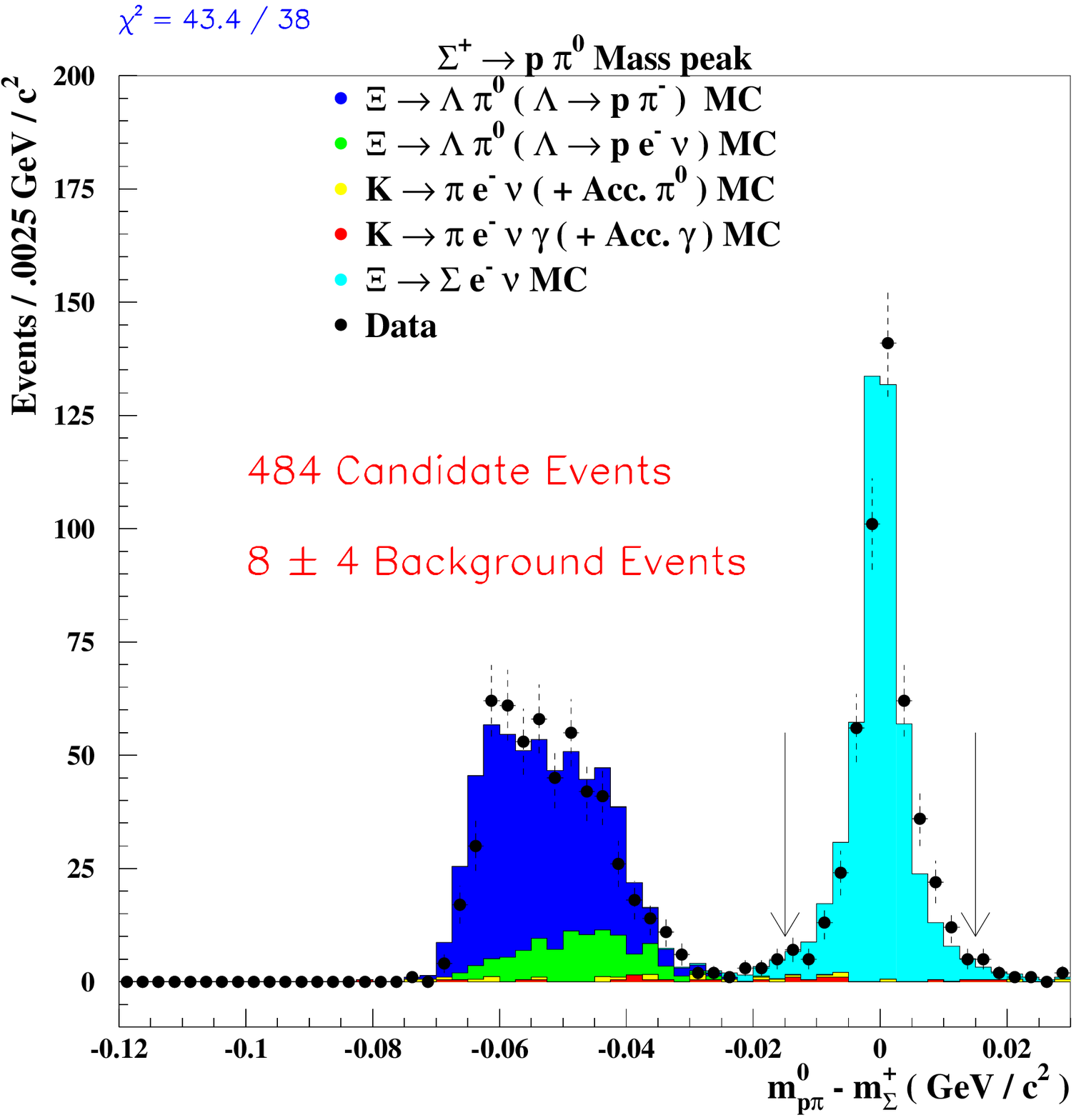,%
height=0.47\linewidth}
}
\mbox{
\epsfig{file=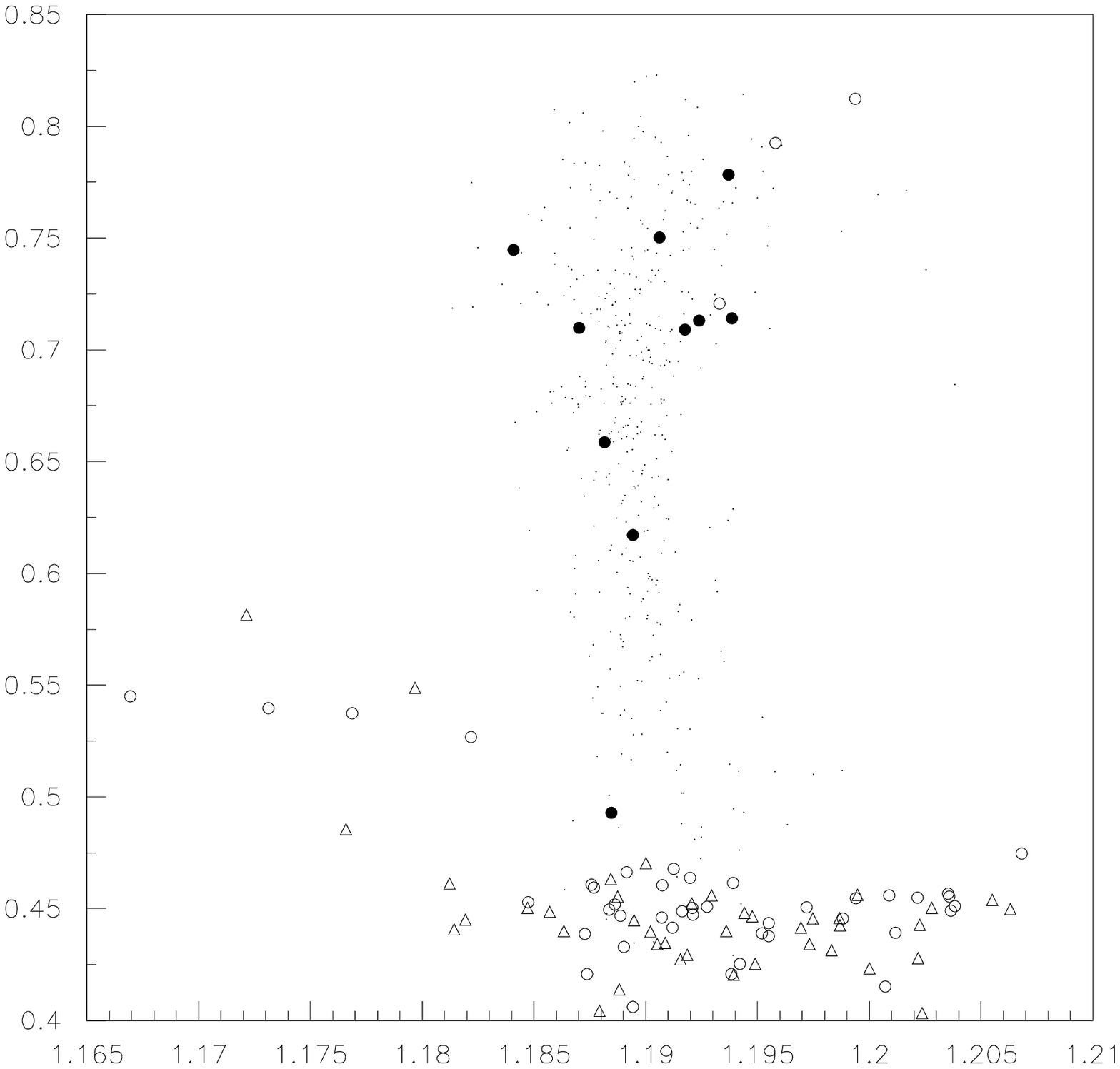,%
height=0.47\linewidth}
}
\end{center}
\caption{A very clean sample of $\Xi^0$ beta decay events from the KTeV
1997 run is shown on the left; these were used for the form-factor
measurements. On the right is the sample of $\Xi^0$ muonic decays from
the KTeV 1999 run, plotting the mass of $p\pi^0$ vs $\pi^+\mu^-\pi^0$;
the smaller dots are Monte-Carlo simulation of $\Xi^0$ muonic decays,
the circles the correct-charge-sign data, and the triangles the
wrong-charge-sign data (anti-hyperon production has a 10x
suppression).} \label{scc}
\end{figure}

\paragraph*{Semi-leptonic hyperon decay physics analyses} accessible in
KTeV
are the beta decay $\Xi^0 \rightarrow \Sigma^+ \hspace{0.1cm} e^{-}
\hspace{0.1cm} \bar{\nu}_{e}$ and muonic decay $\Xi^0 \rightarrow
\Sigma^+ \hspace{0.1cm} \mu^- \hspace{0.1cm} \bar{\nu}_{\mu}$. They are
important to study for their weak decay form-factors which give an
understanding of their underlying structure. In the V-A formulation the
transition amplitude of beta decay is
\begin{equation}
M = \frac{G}{\sqrt{2}} <\Sigma|J^{\lambda}|\Xi>{\bar{u}_{e}}
\gamma_{\lambda} (1 + \gamma_{5}) u_{\nu}
\end{equation}
The V-A hadronic current can be written as
\begin{eqnarray}
<\Sigma|J^{\lambda}|\Xi>\;= {\cal C} \; i\; \bar{u}(\Sigma) & [ &
f_{1}\gamma^{\lambda} + f_{2} \frac{\sigma^{\lambda
\upsilon}\gamma_{\upsilon}}{M_{\Xi}} +
f_{3} q^{\lambda} \frac{M_e}{M_{\Xi}} + \nonumber \\
& [ & g_{1} \gamma^{\lambda} + g_{2} \frac{\sigma^{\lambda \upsilon}
\gamma_{\upsilon}}{M_{\Xi}} + g_{3} q^{\lambda}
\frac{M_e}{M_{\Xi}}\hspace{0.25cm} ] \hspace{0.1cm} \gamma_{5}
\hspace{0.25cm} ]\;u(\Xi)
\end{eqnarray}
where ${\cal C}$ is the CKM matrix element and $q$ is the momentum
transfer. There are 3 vector form-factors: $f_1$ (vector), $f_2$ (weak
magnetism) and $f_3$ (an induced scalar); plus 3 axial-vector
form-factors: $g_1$ (axial vector), $g_2$ (weak electricity) and $g_3$
(an induced pseudo-scalar). All six form-factors are real if $T$-
invariance is valid. The quark model predicts a non-zero but small
$g_2$ form-factor if SU(3) breaking is sizable, but the standard model
assumes this term is zero. Figure \ref{ff} shows the expected changes
in these observable form-factors in the standard model and various
symmetry breaking schemes. The $g_3$ form-factor is expected to be
large ({\it i.e.} $\frac{\textstyle g_3}{\textstyle g_1} \sim 8$), but
it is multiplied by $\frac{\textstyle M_e}{\textstyle M_\Xi}$ making
this term negligibly small so as not to contribute any noticeable
effect. However, for the muonic decay this may no longer be assumed.
Furthermore, neither of these decays had previously been observed, so
measuring their branching ratio was also important as a test of the
standard model, and in the case of the muon decay this could be the
first place to look for a form-factor that substantially depends upon
the mass of the charged lepton. The final results for the beta decay
are a branching ratio of ($2.60 \pm 0.11 \pm 0.16) \times 10^{-4}$,
based on 626 events, where the first error is statistical and the
second systematic, and the theoretical expectation is $2.6 \times
10^{-4}$. For the muonic decay a preliminary branching ratio is ($3.5
^{+2.0}_{-1.0}\ ^{+0.5}_{-1.0}) \times 10^{-6}$ based on 5 events,
while the asymmetric error bars are from the small number of events and
Poisson statistics at the 68\% C.I. \cite{feldman}; theoretically
expected is $2.6 \times 10^{-6}$. A larger sample of these events has
been obtained in the 1999 KTeV run and are shown in figure \ref{scc}
right.

\paragraph*{A very clean sample}
of $\Xi^0$ beta decays, figure \ref{scc} left were obtained by using
the electron identification of the TRD detector. These data were used
to measure the form-factor $g_2$, for which a non-zero value would
indicate new physics beyond the standard model. The decay of the
$\Sigma^+$ has a 98\% analyzing power, and this fact makes it
equivalent to a fully polarized beam. However, spin alignment magnetics
gave the ability to control this, and then to test the technique on the
much larger normal-mode decay $\Xi^0 \rightarrow \Lambda^0 \pi^0$. By
working in the $\Sigma^+$ reference frame, all of the form-factors
could be determined by measuring the angular distribution of the proton
relative to the electron neutrino (we typically use the reconstructed
transverse neutrino direction) (see figure \ref{ff}) or by measuring
the electron energy spectrum. We can also test the technique by
comparing the proton direction relative to the reconstructed $\Xi^0$.
The final four form-factors are: $f_1 = 0.99 \pm 0.14$,
$\frac{\textstyle g_1}{\textstyle f_1}=1.24 \pm 0.27$,
$\frac{\textstyle f_2}{\textstyle f_1}=2.3 \pm 1.3$, and
$\frac{\textstyle g_2}{\textstyle f_1}= -1.4 \pm 2.1$. This analysis
used the previously quoted branching ratio and permitted the $g_2$
form-factor to float. The $g_2$ value is consistent with zero and in
another analysis it was constrained to be zero and the remaining
form-factors re-analyzed; they remained essentially unchanged. For a
more detailed description see \cite{bright}.
\begin{figure}[h]
\begin{center}
\mbox{
\epsfig{file=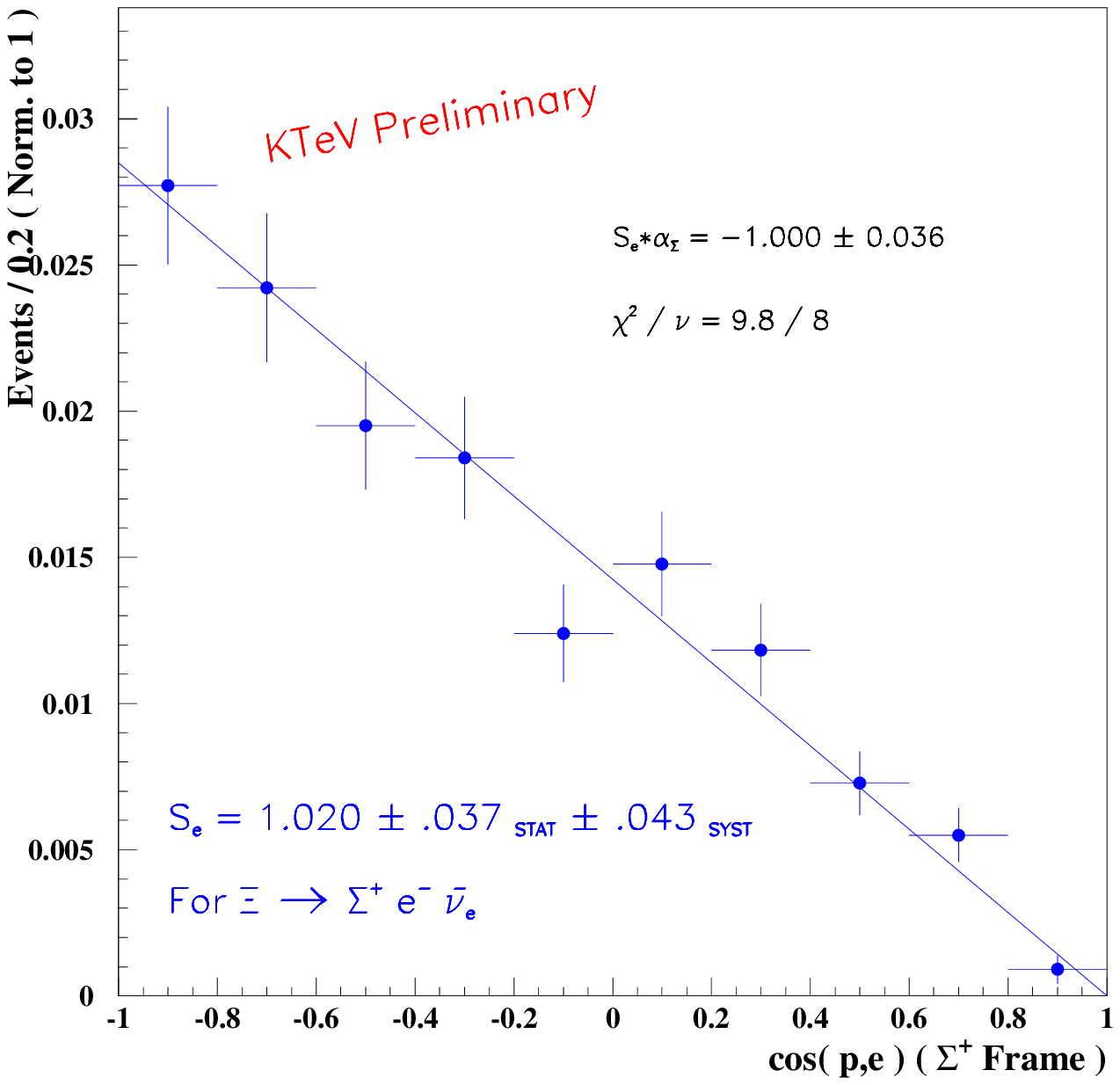,%
height=0.47\linewidth}
}
\mbox{
\epsfig{file=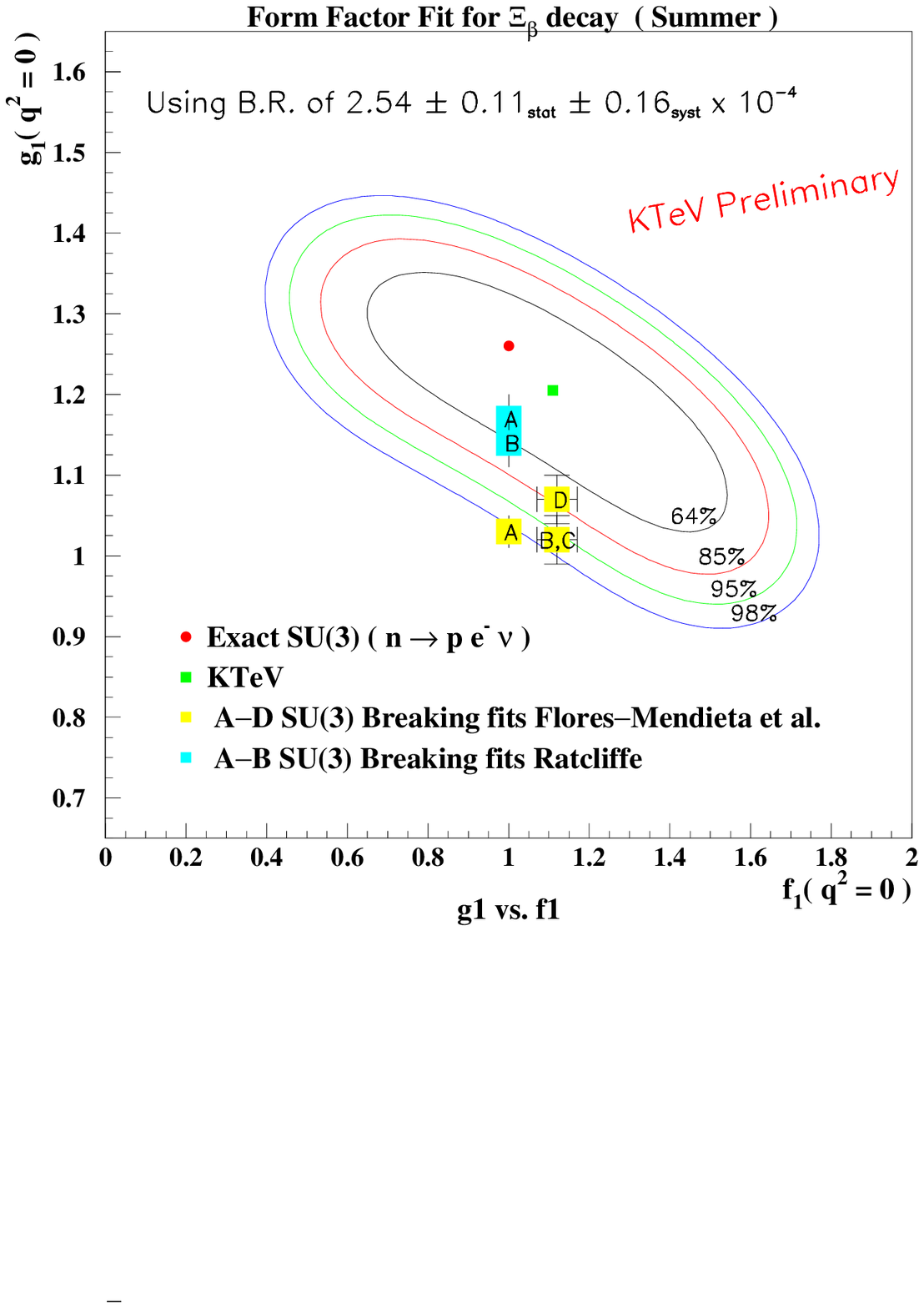,%
height=0.47\linewidth}
}
\end{center}
\caption{On the left is a plot of the $\Xi^0$ beta decay events of the
cosine of the angle between the proton and electron in the center of
mass of the $\Sigma^+$, and on the right is the best fit to the
form-factors $f_1$ and $g_1$.} \label{ff2}
\end{figure}
\begin{figure}[h]
\begin{center}
\mbox{
\epsfig{file=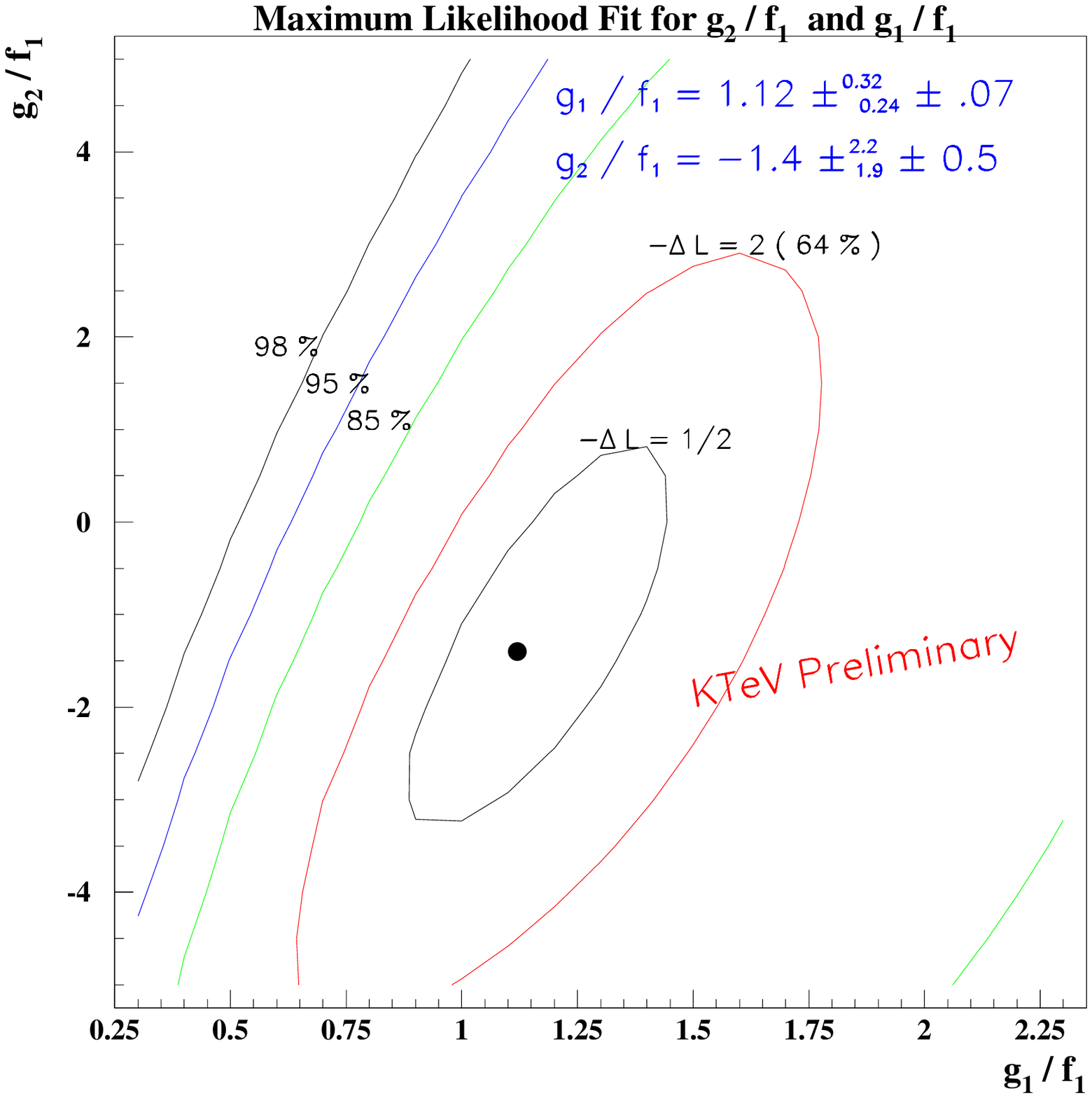,%
height=0.47\linewidth}
}
\end{center}
\caption{A determination of $g_2$ form-factor from the clean $\Xi^0$
beta decay sample is shown along with probability contours.} \label{g2}
\end{figure}

\section{Future Hyperon Beta Decay Measurements}
The following is a list of the most important hyperon beta decay
measurements that should be done and why, and which experiments, with
no or minor modifications, may be able to perform these studies.

\paragraph*{{\em CP} and $T$ violation studies with hyperons:}The subject
of
{\em CP} violation, first seen with the neutral $K^0$ system and hints
now just emerging with the $B^0$ meson, is an important topic to extend
to baryons. The first place this might be able to emerge is with
hyperons that can be produced copiously. The interest in this physics
topic is covered elsewhere in these proceedings \cite{kaplan}. However,
a minor point not covered there is that a large anti-hyperon beta decay
sample could be fertile ground in which to compare the branching ratio
with that of regular-matter hyperon beta decay.

\begin{figure}[h]
\begin{center}
\mbox{
\epsfig{file=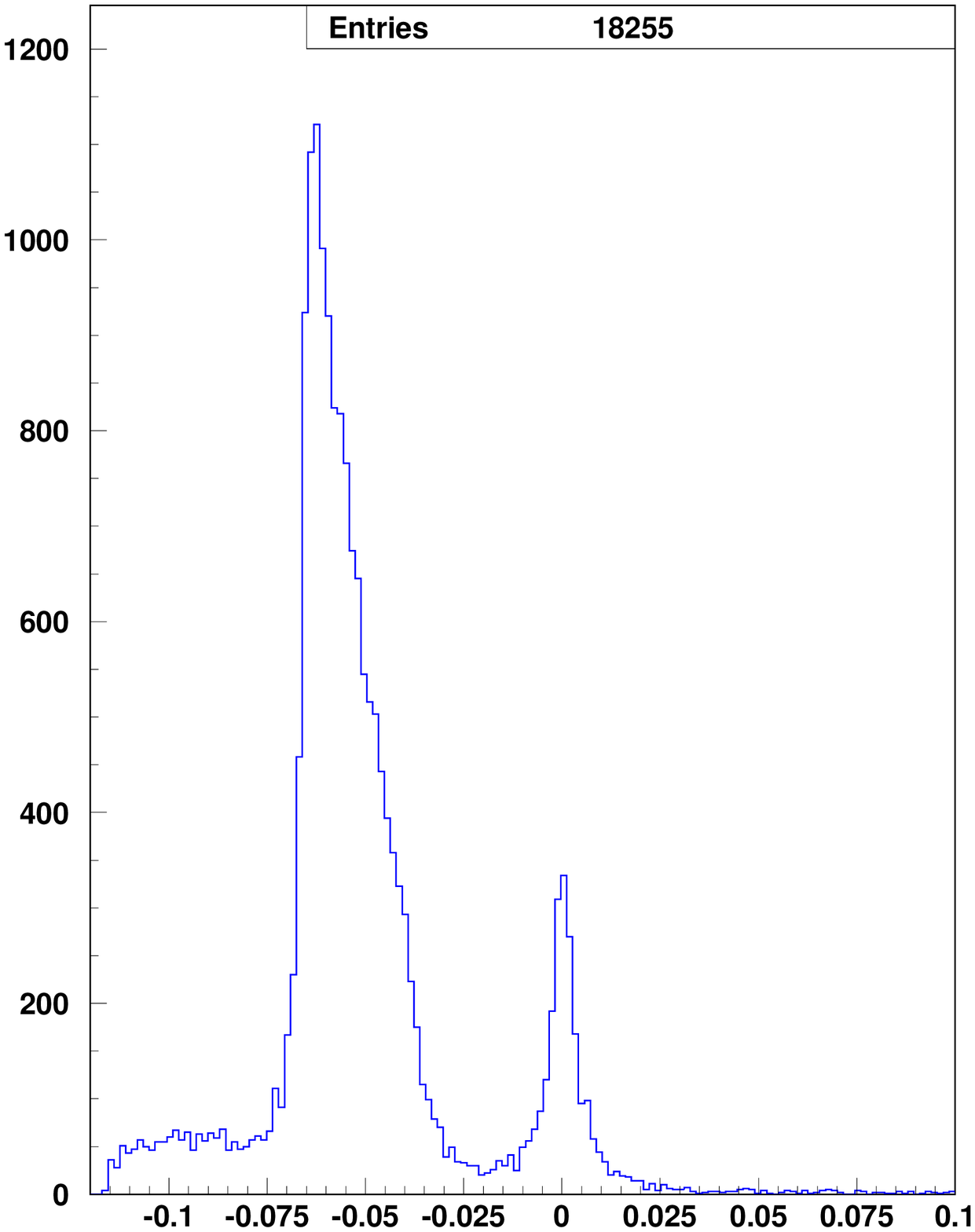,%
height=0.6\linewidth}
}
\end{center}
\caption{The KTeV 1999 run has an additional 2100 $\Xi^0$ beta
decay events.}
\label{ktev99}
\end{figure}

\paragraph*{High-statistics sample of hyperon beta decays:}
This would permit a precision form-factor measurement which is
important as a test of the standard model as well as a good means of
searching for new physics not currently in the standard model. The KTeV
result mentioned in section 1 from the 1997 run is the start of such a
measurement because it will permit the best form-factor analysis with
the $\Sigma^+$ self-analyzing power. There is a three times larger
sample from the 1999 run (see figure \ref{ktev99}) which, when merged,
will offer a great improvement in the form-factor analysis. However,
there is the potential for a ten-fold improvement over the full KTeV
hyperon sample (1997 plus 1999 data) with a dedicated run using the
$K_{\rm short}$ target at the NA48 experiment at CERN scheduled for
2002 \cite{addem2}.

The term $V_{us}$ as measured with $Ke3$ decays, $K^0 \rightarrow \pi^+
e^- \bar{\nu}$, and those from three hyperon beta decays do not have
perfect agreement, see table 1. In principle $V_{us}$ measured from
these decays should be the same, but what is actually being measured is
$|f_1 V_{us}|$. However, no particle, neither meson nor baryon, has
free quarks to measure $V_{us}$ directly. It is presumed that this
experimentally observed discrepancy is due to the strong force
potential that the quarks are in, hence the implication that the
measurement with the mesons might be closer to reality, but even this
is a poor approximation.

\begin{table}[h]
\caption{$V_{us}$ as determined by various hyperon beta decays, and
from $Ke3$ meson decay.}
\begin{center}
\begin{tabular}{|l|c|c|}
\hline
Decay & $V_{us}$ & Uncertainty \\
\hline
$K^0_L \rightarrow \pi^+ e^-\bar{\nu}$ & 0.2188 & $\pm$0.0016 \\
\hline
$\Lambda^0 \rightarrow p\;e^- \bar{\nu}$ & 0.2130 & $\pm$0.0020 \\
$\Sigma^- \rightarrow n\;e^- \bar{\nu}$  & 0.2318 & $\pm$0.0040 \\
$\Xi^- \rightarrow \Lambda^0 e^- \bar{\nu}$ & 0.2434 & $\pm$0.0068 \\
\hline
\end{tabular}
\end{center}
\label{vus}
\end{table}

While seeing $g_2 \ne 0$ would be an indication of new physics beyond
the standard model, it is difficult to observe with any present
experiment. Hence the need for a dedicated experiment. It has also been
noted that hints for a non-zero $g_2$ form-factor may already exist,
because when $\frac{\textstyle g_2}{\textstyle g_1} \equiv 0.2$ then
all of the experimental measurements of $V_{us}$ using hyperon beta
decays (see table 1) come out equal to $0.220 \pm 0.004$, in agreement
with the $Ke3$ determination \cite{donoghue}. However, there are other
possible explanations to account for the discrepancy. Obviously another
measurement from a fourth hyperon beta decay would be useful, as would
a high-statistics measurement of any one hyperon beta decay, the best
being the $\Xi^0$ beta decay because of the $\Sigma^+$ analyzing power.

\paragraph*{$\Lambda^0-\Sigma^0$ mixing:}
It is known that the mesons experience mixing between neutral states,
and a similar mixing with $\Lambda-\Sigma^0$ is expected \cite{Karl}.
This can easily be tested by measuring any difference between the
branching ratio of the $\Sigma^{\pm}$ beta decays $\Sigma^+ \rightarrow
\Lambda^0 e^+ \nu$ and $\Sigma^- \rightarrow \Lambda^0 e^- \bar{\nu}$.
Both of these decays are badly measured and a 2\% branching ratio
measurement would suffice.

Furthermore, the $\Lambda^0$ has a 64\% analyzing power, although not
98\% like the $\Sigma^+ \rightarrow p \pi^0$ decay used in $\Xi^0$ beta
decay, a large sample of either one of these decays could help resolve
the $V_{us}$ discrepancy between the hyperon beta decays and that from
$Ke3$ decays. It has also been pointed out that these $\Sigma^{\pm}$
beta decays could place the best limit on SU(3) symmetry breaking since
in these decays $V_{us}$ does not enter since they have just an
axial-current term.

\paragraph*{Form-factors outside of the normal octet:}
Measurements of anti-hyperon beta decays would give another test of
$V_{us}$ that may explain the discrepancy with $V_{us}$ measured using
mesons, which are quark anti-quark states, and this may be contributing
to the discrepancy with that of hyperons. A measurement of this would
also be a test of {\em CP} and $T$ violation in hyperons just by
comparing branching ratios at the 0.1\% level, but the real interest is
in the form factor similarity for anti-matter, which has never been
tested.

Measurements of the form-factors outside of the octet, such as with the
$\Omega^- \rightarrow \Xi^0 e^- \bar{\nu}$, or with charm-strange
baryons, $\Lambda^+_c \rightarrow \Lambda^0 e^+ \nu$, would give
another measurements of $V_{us}$ and the first measurement of $V_{cs}$
with baryons. The $\Omega^-$ beta decay could also be compared to the
matrix elements predicted by SU(6). Hyperon mixing of
$\Sigma^0-\Lambda$ could also be seen here by observing its decay into
$\Sigma^0 e^- \bar{\nu}$.

\paragraph*{Muonic hyperon decays:}
The hyperon muonic decays all have poorly measured branching ratios,
and we have never had a large enough sample to be used in a form-factor
measurement. This could be useful in several ways. First, a tagged
decay such as $\Omega^- \rightarrow \Lambda^0 K^-$ where $K^-
\rightarrow \pi^- \pi^+ \pi^-$ and $\Lambda^0 \rightarrow p \mu^-
\bar{\nu}$ could be of great assistance. This would help because the
presence of a $K^-$ decaying into three charged pions would indicate
unambiguously the presence of a $\Lambda^0$, and an experiment such as
HyperCP (E871) at Fermilab has excellent muon identification to
distinguish this. Other hyperon and kaon backgrounds would be
eliminated leaving only the charged-pion decay to $\mu^- \bar{\nu}$ to
contend with. For a branching-ratio measurement this contribution can
be simulated, and if it can be eliminated by topology constraints, then
a muonic decay form-factor can be extracted. The advantage of the
$\Xi^0$ muonic decay is that there are no competing two-body
backgrounds that contain a $\pi^-$, hence no background from this
source. As can be seen from the cleanliness of the KTeV $\Xi^0$ muonic
decay, see figure \ref{scc} right, the signal is exceptionally strong
and well separated from the kaon backgrounds.

The importance of measuring hyperon muonic decays is that it is the
only process where the $g_3$ form factor is expected to contribute any
noticeable charged-lepton mass effects \cite{linke}. Although small,
$\sim 15$\%, radiative corrections are expected to be half of this
value $\sim 7$\%. Nevertheless, a sample of 300 to 500 such events is
expected from the NA48 special run \cite{addem2}. So there is a future
experimental possibility with this type of decay. Another decay that is
expected to have a lot of hyperon muonic decays is the $\Omega^-$
system. Here, due to the large Q value of the beta and muonic decays,
the branching ratio for $\Omega^-$ is high: $\sim1 \times 10^{-3}$!
Maybe these can be extracted cleanly from the HyperCP experimental data
sample for improved branching ratio measurements. Due to the high Q
(released energy) phase space would not restrict their branching ratio,
so just a comparison of the beta and muonic decay branching ratio in
this system is a first test of the form factor equivalence. A clean
sample could yield an independent form-factor measurement where $g_3$
is large enough to be seen, or rule out its existence.

\section*{Conclusions}
Beta decays have been a source of great physics discoveries, since the
prediction of the neutrino to account for an anomalous electron energy
spectrum. With hyperons they allow an independent measurement of
$V_{us}$ in the standard model, and are a great place to hunt for
physics beyond the standard model. They may even hold some clues to the
unification of the strong nuclear force with electro-weak theory if the
form-factor $g_2$ can be explicitly shown to be non-zero. This is
because although they are from a weak decay, the strong force has a
substantial role in the hyperons themselves. When a more massive
charged lepton such as the muon replaces the electron of beta decay,
this is the only place that can show the effect of the $g_3$
form-factor. All of these exciting topics makes for continued interest
in studying hyperon beta decays.

I would like to thank the organizers, D. Kaplan and H. Rubin,
for the opportunity to
present my perspective on future hyperon beta-decay experiments. I also
wish to thank J. Rosner for many useful discussions.

\end{document}